%% file: BIOM2019314Marx_v2.tex
\documentclass[useAMS,referee,usenatbib]{biom}
\usepackage{amsmath,amsfonts,enumerate}
\usepackage[hidelinks]{hyperref}
\usepackage{graphicx}

\usepackage{xr}
\input{bold.tex}

\newcommand{\msd}{\mbox{MSD}}
\newcommand{\tr}{\mbox{tr}}
\newcommand{\vecop}{\mbox{vec}}

\title[Distance-Based Reliability]{Generalized reliability based on distances}

\author{Meng Xu$^{1,*}$\email{mxu@campus.haifa.ac.il}, 
Philip T.\ Reiss$^{1,**}$\email{reiss@stat.haifa.ac.il}, and 
Ivor Cribben$^{2,***}$\email{cribben@ualberta.ca} \\
$^{1}$Department of Statistics, University of Haifa, Haifa 31905, Israel \\
$^{2}$Department of Finance and Statistical Analysis, Alberta School of Business,\\
University of Alberta,
Edmonton, AB T6G 2R6, Canada}

\begin{document}


\date{{\it Revised Jan} 2020.}



\pagerange{\pageref{firstpage}--\pageref{lastpage}} 
\volume{00}
\pubyear{2020}




\label{firstpage}


\begin{abstract}
The intraclass correlation coefficient (ICC) is a classical index of measurement reliability. With the advent of new and complex types of data for which the ICC is not defined, there is a need for new ways to assess   reliability. To meet this need, we propose a new distance-based intraclass correlation coefficient (dbICC), defined in terms of arbitrary distances among observations.  We introduce a bias correction to improve the coverage of bootstrap confidence intervals for the dbICC, and demonstrate its efficacy via simulation. We illustrate the proposed method by analyzing the test-retest reliability  of brain connectivity matrices derived from a set of repeated functional magnetic resonance imaging scans. The Spearman-Brown formula, which shows how more intensive measurement increases reliability, is extended to encompass the dbICC. \end{abstract}

%



\maketitle


%

\section{Introduction}
\label{s:intro}
With  the increasing availability of new and complex forms of data, there is a corresponding need for new ways to assess measurement reliability. This article aims to help meet this need by reformulating the intraclass correlation coefficient (ICC), a standard index of reliability, in terms of distances between observations.

We begin by defining the ICC as developed in classical test theory \citep{lord1968,fleiss1986,mair2018}, which views a measured scalar quantity $X$ as the sum of an underlying true score $T$ and an error term $E$. Suppose we have a sample of $I$ individuals with true real-valued scores $T_1,\ldots,T_I$ drawn from a population with variance $\sigma^2_T$; and that for each $i$, the $i$th individual is measured $J_i$ times, yielding observations
\begin{equation}\label{xte}
	X_{ij}=T_i+\varepsilon_{ij},
\end{equation}
$j=1,\ldots,J_i$, where the $\varepsilon_{ij}$'s are drawn from a distribution with mean 0 and variance $\sigma^2_{\varepsilon}$, independently of each other and of the $T_i$'s. Then for distinct $j_1,j_2\in\{1,\ldots,J_i\}$, the correlation between the $j_1$th and $j_2$th observations for individual $i$ is easily shown to be 
\begin{equation}\label{eq:classICC}
	\rho=\frac{\sigma^2_T}{\sigma^2_T+\sigma^2_{\varepsilon}}.
\end{equation}
This quantity is the classical ICC.

Reliability measures for more complex settings include replacing model \eqref{xte} with the generalizability theory model of \cite{cranford2006procedure},    as well as generalizations of \eqref{eq:classICC} to multivariate data \citep{alonso2010}, including high-dimensional data
\citep{shou2013}.  All of these extensions assume a model that is more complex than \eqref{xte}, but still of an additive (signal plus noise) form. However, for complex objects that are measured or estimated in modern biomedical research, such as  motion patterns or brain networks, such an additive representation is typically inapplicable. There is thus a need for a new reliability index appropriate for general data objects.

\begin{figure}[!h]
	\centering
	\includegraphics[width=.8\textwidth]{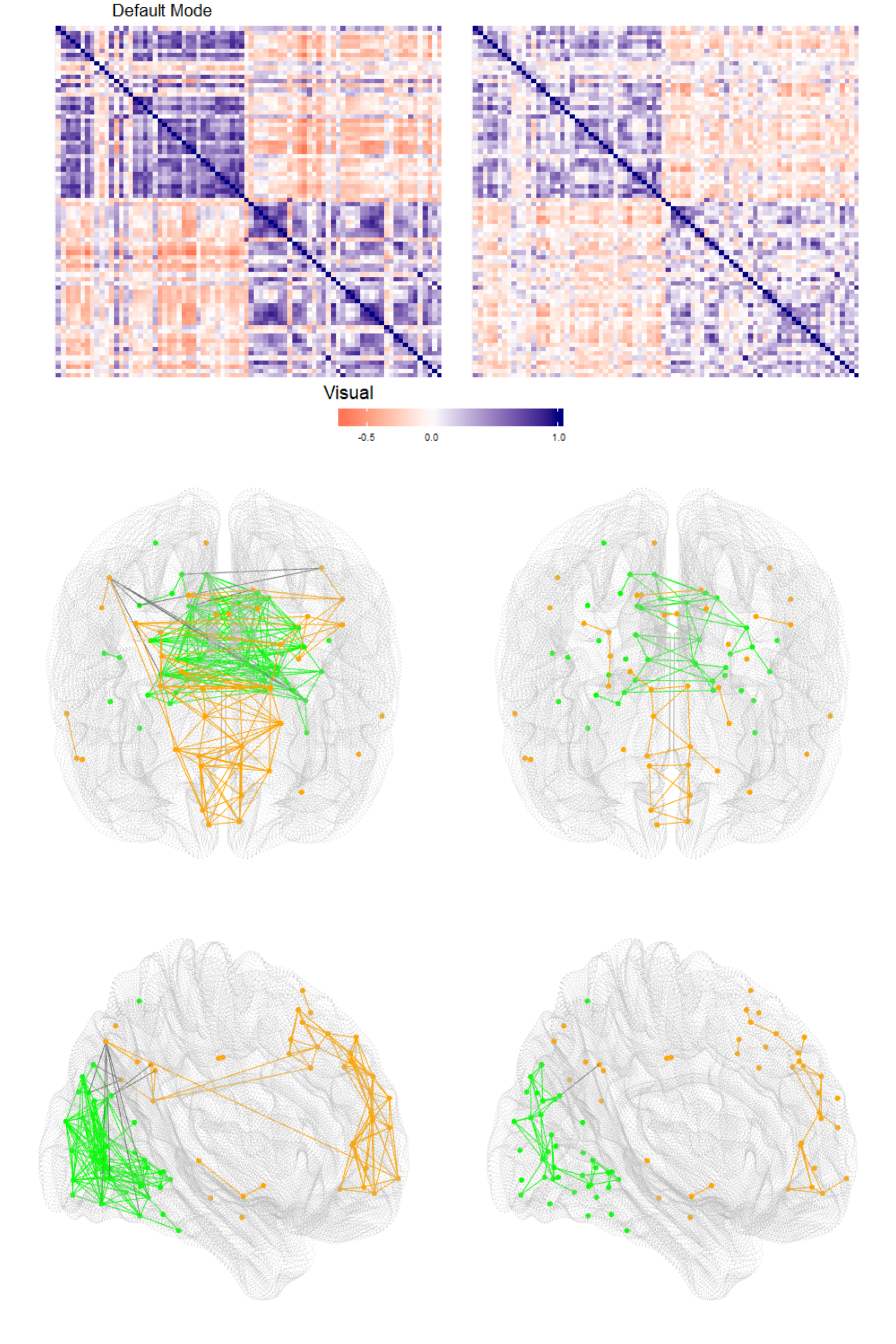}
	\caption{\emph{Top:} 	Matrices $\bR$ of correlations among 80 ROIs comprising the default mode network and visual network  in our fMRI data set.  The left and right matrices, respectively,  attain the highest and lowest connectivity scores $-\log|\bR|$ observed in our data set. \emph{Middle:} Brain maps (axial view) corresponding to the same two correlation matrices, and displaying pairs of regions with absolute correlation above 0.6. Orange nodes and links refer to the default mode network; green nodes and links refer to the visual network;  links  between the two networks are shown in black.  \emph{Bottom:} Same brain maps, sagittal view. The fMRI data are presented in Section~\ref{funco}, and the connectivity score $-\log|\bR|$ is discussed briefly in Web Appendix~\ref{supp-logR}.}
	\label{brains}
\end{figure}

Our work was motivated by the study of functional connectivity in the human brain by means of resting-state functional magnetic resonance imaging (fMRI). Briefly, fMRI produces a time series of brain activity, known as the blood oxygen level dependent (BOLD) signal, at each of a set of regions of interest (ROIs).  \emph{Resting-state} fMRI means that the participants in the study were not performing any particular task or viewing a stimulus during the brain scan. Functional connectivity  refers to association among activity levels in different parts of the brain, and can be measured in many ways \citep{yan2013}. One of the most common functional connectivity measures is a simple Pearson correlation matrix of regional BOLD signals. Figure~\ref{brains} displays two such correlation matrices,  along with associated brain graphs, for a set of 80 ROIs to be discussed in Section~\ref{funco}. These particular examples were chosen to illustrate high and low connectivity, according to a metric described in Web Appendix~\ref{supp-logR}.
 In order to be confident that such correlation matrices, and the scientific conclusions derived from them, are trustworthy and reproducible, it is necessary first to be able to assess their reliability. Our proposed methodology offers a means to that end. 

Our basic proposal, a reformulation of the ICC based on distances between observations, is outlined in Section~\ref{dbrm}, and estimation of the resulting  reliability index is discussed in Section~\ref{estit}. An application to an fMRI data set is presented in Section~\ref{funco}. In Sections~\ref{gensbf}--\ref{mgd} we extend the Spearman-Brown formula, a fundamental result in reliability theory, to our distance-based ICC, and revisit our fMRI data set in light of this extension. A concluding discussion appears in Section~\ref{s:discuss}.

\section{Distance-based reliability measurement} \label{dbrm}

A novel reliability index applicable to general data objects can be defined by re-deriving the ICC \eqref{eq:classICC} in terms of squared distances among observations.
Let $\msd_b=E_{i_1\neq i_2}[(X_{i_1j_1}-X_{i_2j_2})^2]$ and $\msd_w=E_{j_1\neq j_2}[(X_{ij_1}-X_{ij_2})^2]$ be the mean squared differences for measurements between and within individuals, respectively. Then $\msd_b=2\sigma_T^2+2\sigma_{\varepsilon}^2$ and $\msd_w=2\sigma_{\varepsilon}^2$, and thus the ICC \eqref{eq:classICC} can be re-expressed as 
\begin{equation}\label{dbicc}\rho=1-\frac{\msd_w}{\msd_b}.\end{equation}
The advantage of expression \eqref{dbicc} is that, unlike \eqref{eq:classICC}, it extends straightforwardly to general data objects (curves, networks, etc.), as long as a distance or dissimilarity $d(\cdot,\cdot)$ between such objects is defined. One simply redefines $\msd_b$, $\msd_w$ in \eqref{dbicc} in a more general sense, as the between- and within-individual mean squared \emph{distances} \begin{equation}\label{msdbw}\msd_b=E_{i_1\neq i_2}[d(X_{i_1j_1},X_{i_2j_2})^2]\quad\mbox{and}\quad\msd_w=E_{j_1\neq j_2}[d(X_{ij_1},X_{ij_2})^2].\end{equation} 
Henceforth we shall refer to \eqref{dbicc}, with $\msd_b,\msd_w$ given by \eqref{msdbw}, as the \emph{distance-based intraclass correlation coefficient}, or dbICC. 

We note that the same general strategy, of re-deriving variance-based formulas  in terms of sums of squared distances, has been used previously to formulate distance-based hypothesis tests \citep{mcardle2001,mielke2007,reiss2010}.

A simple example of extending  \eqref{xte} beyond the scalar real-valued case is to let $T_i,\varepsilon_{ij}$ be mutually independent random \emph{vectors}, with covariance matrices $\bSigma_T,\bSigma_{\varepsilon}$ respectively, and let $d$ be the Euclidean distance. Then \eqref{dbicc} reduces straightforwardly to
\begin{equation}\label{iicc}\rho=1-\frac{\tr(\bSigma_{\varepsilon})}{\tr(\bSigma_T+\bSigma_{\varepsilon})}=\frac{\tr(\bSigma_T)}{\tr(\bSigma_T+\bSigma_{\varepsilon})},\end{equation}
 the multivariate reliability measure referred to as $R_T$ \citep{alonso2010}, and as I2C2 \citep{shou2013} for images viewed as vectors. Thus the dbICC is an extension of these measures to more general distances and data types.

\section{Estimating the dbICC}\label{estit}

\subsection{Point estimation}
\begin{figure}\centering
\includegraphics[width=\textwidth]{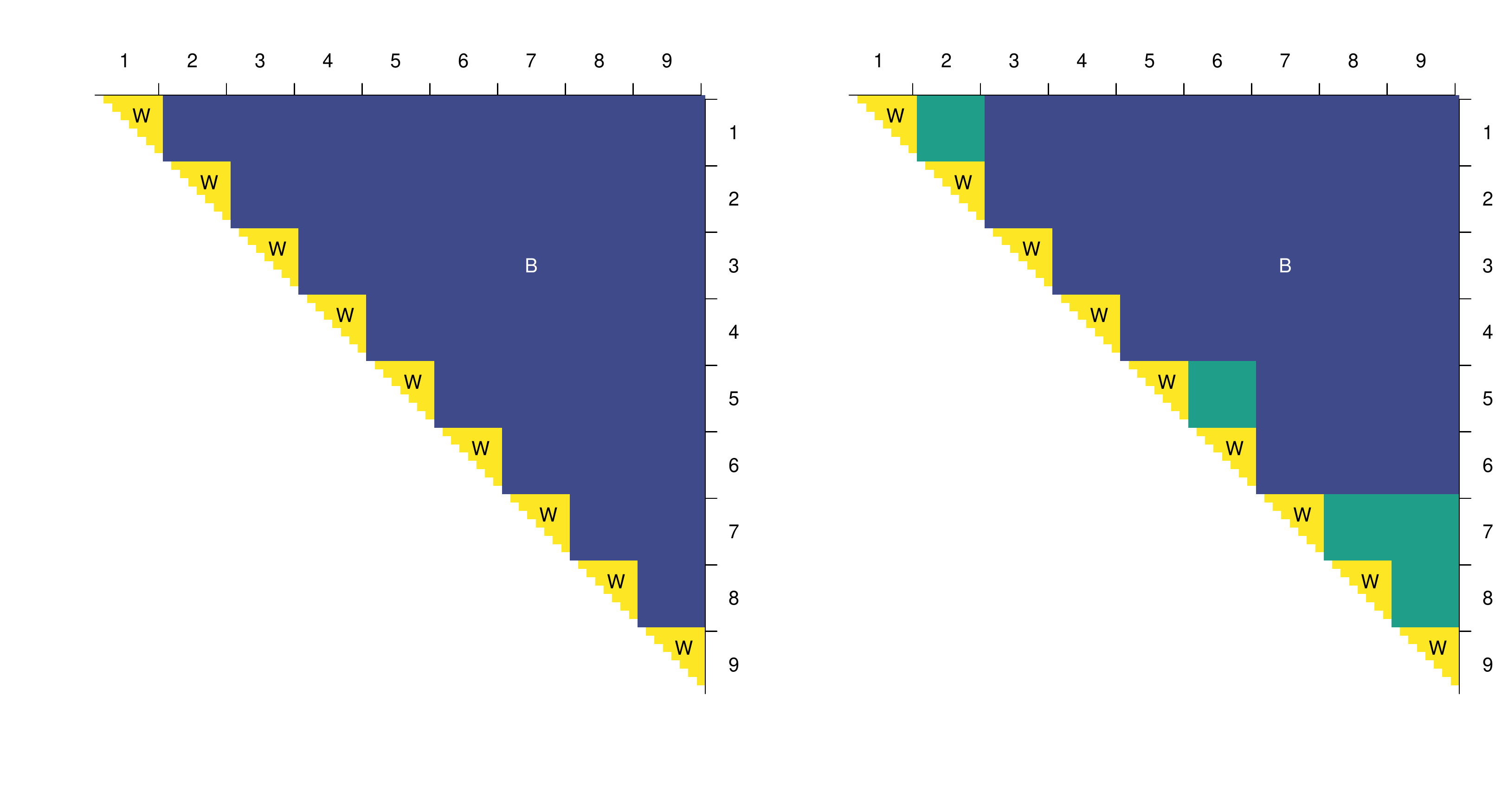}\caption{\emph{Left:} Schematic diagram of a matrix of distances among repeated observations of nine individuals, with rows and columns grouped by individual. Distances in the half-squares along the diagonal are within-individual (W), while the rest are between-individual (B). \emph{Right:} A similar diagram, but for a bootstrap sample with repeated observations. Distances shown in green are nominally between-individual, but in reality they are within-individual.}
\label{bwm}
\end{figure}

Like the classical ICC \eqref{eq:classICC}, the proposed dbICC \eqref{dbicc} can be estimated in practice by plugging in consistent estimates of the population quantities \eqref{msdbw}, as follows:
\begin{equation}\label{dbicchat}\hat{\rho}=1-\frac{\widehat{\msd}_w}{\widehat{\msd}_b}\end{equation}
where
\begin{equation}\label{msdb}\widehat{\msd}_b=\frac{\sum_{1\leq i_1<i_2\leq I}\sum_{j_1=1}^{J_{i_1}}\sum_{j_2=1}^{J_{i_2}}d(X_{i_1j_1},X_{i_2j_2})^2}{\sum_{1\leq i_1<i_2\leq I}J_{i_1}J_{i_2}},\end{equation}
\begin{equation}\label{msdw}\widehat{\msd}_w=\frac{\sum_{i=1}^I\sum_{1\leq j_1<j_2\leq J_i}d(X_{ij_1},X_{ij_2})^2}{\sum_{i=1}^I {J_i \choose 2}}.\end{equation}
 Figure~\ref{bwm}  illustrates this schematically for a distance matrix with rows and columns grouped by individuals: one estimates $\mbox{MSD}_b,\mbox{MSD}_w$ by averaging the between- and within-individual distances (B and W), respectively. 

\subsection{Bootstrap confidence intervals}
The dbICC is intended for distance functions whose distribution may not be known. It is thus natural to turn to nonparametric bootstrapping as a distribution-free approach to interval estimation for the dbICC. For $r=1,\ldots,B$ with suitably large $B$, let $\pi_1^r,\ldots,\pi_I^r$ be a sample with replacement from $\{1,\ldots,I\}$; then the $r$th bootstrap sample consists of $X^r_{ij}\equiv X_{\pi_i^rj}$ for $i=1,\ldots,I$ and $j=1,\ldots,J_{\pi_i^r}$. 
  The resulting ICC estimate is 
\begin{equation}\label{rhor}\hat{\rho}^r=1-\frac{\widehat{\msd}^r_w}{\widehat{\msd}^r_b},\end{equation}
where $\widehat{\msd}^r_w,\widehat{\msd}^r_b$ are bootstrap analogues of \eqref{msdb}, \eqref{msdw}: 
\begin{equation}\label{msdbr}\widehat{\msd}^r_b=\frac{\sum_{1\leq i_1<i_2\leq I}\sum_{j_1=1}^{J_{\pi^r_{i_1}}}\sum_{j_2=1}^{J_{\pi^r_{i_2}}}d(X^r_{i_1j_1},X^r_{i_2j_2})^2}{\sum_{1\leq i_1<i_2\leq I}J_{\pi^r_{i_1}}J_{\pi^r_{i_2}}},\end{equation}
\[\widehat{\msd}^r_w=\frac{\sum_{i=1}^I\sum_{1\leq j_1<j_2\leq J_{\pi^r_i}}d(X^r_{ij_1},X^r_{ij_2})^2}{\sum_{i=1}^I {J_{\pi^r_i} \choose 2}}.\]
The interval from the $\alpha/2$ to the $1-\alpha/2$ quantile of the $\hat{\rho}^r$'s can then be used as a $100(1-\alpha)$\% confidence interval.

These bootstrap estimates $\hat{\rho}^r$, however, suffer from   negative bias  \citep[over and above the well-known negative bias of the classical ICC;][]{atenafu2012}. Returning to the example in Figure~\ref{bwm}, consider a bootstrap sample in which individuals 1 and 2 are duplicates, as are individuals 5 and 6 and individuals 7, 8 and 9. Then the blocks shown in the right subfigure in green nominally refer to between-individual differences, but in fact represent within-individual differences. Assuming $\msd_w<\msd_b$, counting these entries as between-individual will tend to result in underestimation of $\msd_b$ and hence in negative bias in \eqref{rhor}. The diagonal entries of these blocks are zero, thereby compounding the bias. To remove this bias, we can simply exclude such blocks from the summations in \eqref{msdbr}; formally, we replace each occurrence of $\sum_{1\leq i_1<i_2\leq I}$ with $\sum_{1\leq i_1<i_2\leq I, \pi^r_{i_1}\neq \pi^r_{i_2}}$.

\subsection{A simulation study}
\begin{figure}
	\centerline{
	\includegraphics[width=\textwidth]{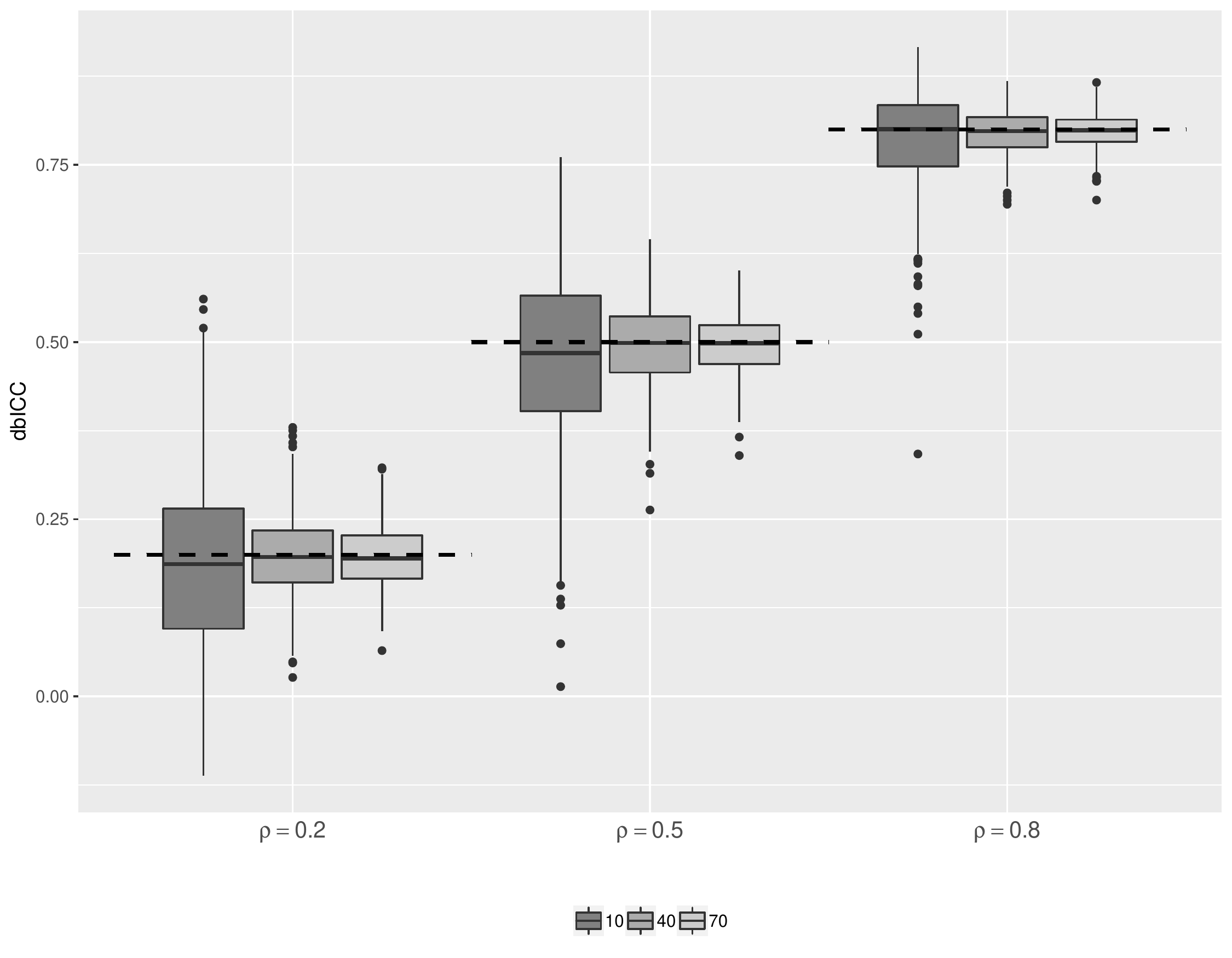}}
	\caption{Boxplots of point estimates of dbICC, for true values $\rho=0.2, 0.5, 0.8$ (indicated by dashed lines) and for $I=10,40,70$.}
	\label{fig:scatter-dbICC}
\end{figure}

Using multivariate data with Euclidean distance (the example from the end of Section~\ref{dbrm}), we conducted a simulation study to assess the accuracy of our point and interval estimates of the dbICC. 
Values $X_{ij}\in \mathbb{R}^2$ were drawn from
\eqref{xte} where  $T_i\sim N_2(0,\bI_2)$ and $\varepsilon_{ij}\sim N_2(0,c\bI_2)$ with $c=4,1,0.25$.
By \eqref{iicc}, the (population) dbICC is then $\rho=\frac{1}{c+1}$, which equals 0.2, 0.5 and 0.8 for the above three values of $c$. The number of subjects $I$ was set to 10, 40 and 70, and the number of measurements per subject $J_i$ fixed at 4. We took 500 replicates with each combination of the above values of $\rho$ and $I$. Boxplots of the dbICC estimates are displayed in Figure~\ref{fig:scatter-dbICC}. The classical negative bias of ICC estimates \citep{atenafu2012} is noticeable for $I=10$ when $\rho=0.2,0.5$, but not for the other settings.

\begin{figure}[!h]
	\centering
	\includegraphics[width=\textwidth]{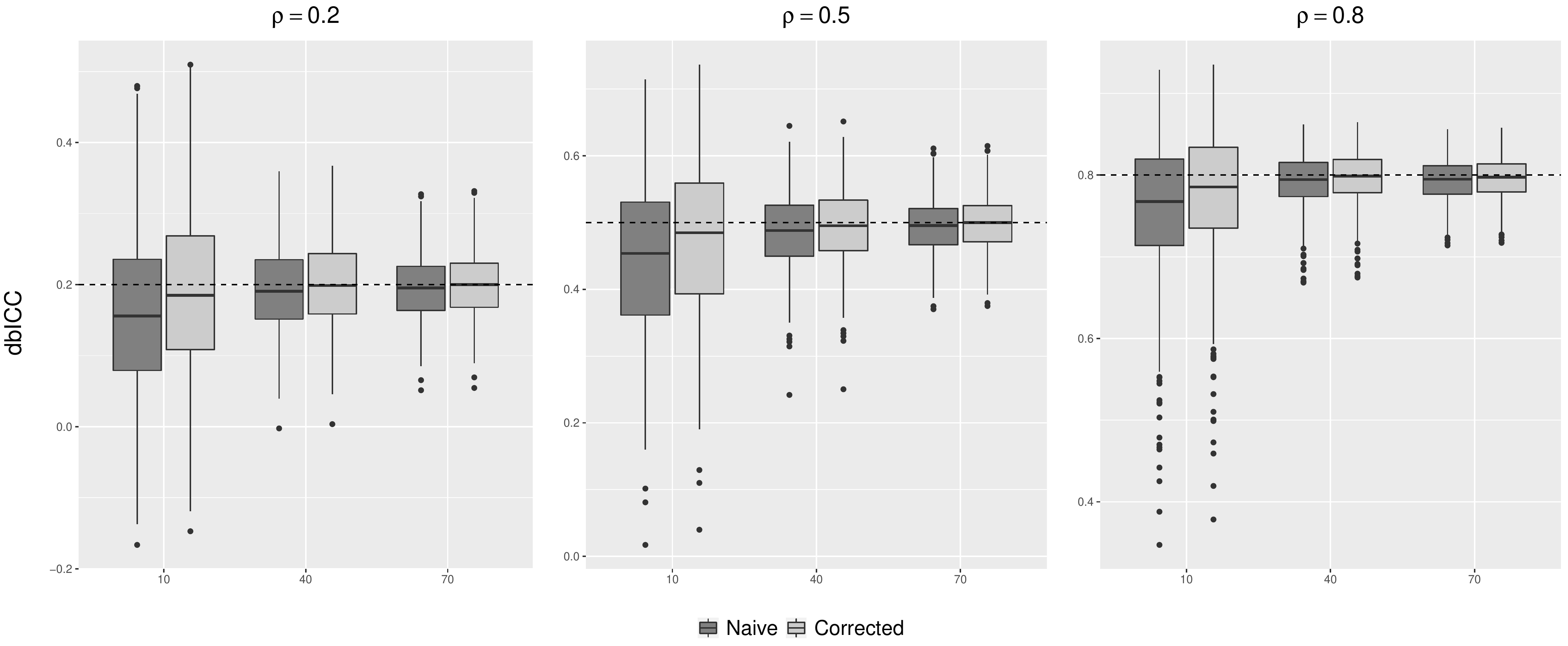}
	\caption{Boxplots of median bootstrap estimate of dbICC, for true values $\rho=0.2, 0.5, 0.8$ (indicated by dashed lines) and for $I=10,40,70$.}
	\label{fig:bootmedian}
\end{figure}

Next we considered bootstrap confidence intervals, with $B=1200$, without and with the bias correction of the previous subsection. 
We performed 500 replicates for each combination of the same $\rho$ and $I$ values as above, again with  $J_i$ fixed at 4. Boxplots of the median of the 1200 bootstrap estimates within each replicate are presented in Figure~\ref{fig:bootmedian}. For $I=10$ and to some extent for $I=40$, the correction yields a marked reduction in the observed negative bias. Accordingly, the coverage of 95\% confidence intervals is improved by the correction, as can be seen in Table~\ref{tb:bootR2}. As noted above, however, a small-sample negative bias (unrelated to bootstrapping) occurs for point estimates of dbICC as for the classical ICC, and hence the coverage remains quite poor for $I=10$.
\begin{table}
	\begin{tabular}{r|rr|rr|rr}
		\hline
		& \multicolumn{2}{c|}{$I=10$} &  \multicolumn{2}{c|}{$I=40$}  &  \multicolumn{2}{c}{$I=70$}  \\ \hline
			 & N & C & N & C & N & C\\
		\hline
		$\rho=0.2$ &86.0 & 90.8 & 91.6 & 93.2 & 92.2 & 92.6 \\ 
		$\rho=0.5$ &84.8 & 90.6 & 91.4 & 92.0 & 94.0 & 94.6 \\
		$\rho=0.8$ &85.2 & 89.6 & 90.6 & 92.6 & 92.8 & 94.2 \\
		\hline\\\\
	\end{tabular}
	\caption{Percent coverage of bootstrap 95\% confidence intervals, na\"ive (N) and corrected (C).} 	\label{tb:bootR2}
\end{table}


\section{Functional connectivity in the human brain}\label{funco}
As noted in the introduction, the dbICC was originally conceived as a way to evaluate the reliability of  functional connectivity measures. To demonstrate how dbICC can be so applied, here we re-examine part of a data set presented by \cite{shehzad2009} in an early study of the test-retest reliability of resting-state functional connectivity.
These authors, followed by others \citep[e.g.,][]{somandepalli2015, choe2017}, focused on ordinary ICC at each of a set of brain locations or connections. The dbICC, by contrast, offers an overall index of reliability for fMRI-based correlation matrices, viewed as \emph{gestalt} measures of functional connectivity.

The data include BOLD time series of length 197,  within each of 333 ROIs derived by \cite{gordon2016}, for  $I=25$ individuals, with $J=2$ such fMRI scans per individual; further details are provided in the Appendix. 
We then computed the distance between each pair of matrices $\bR_1,\bR_2$ among the $25\times2=50$ correlation matrices thus derived, using each of three distance measures:
\begin{enumerate}[(i)]
\item The $\ell_2$ distance (square root of sum of squared differences) between $\vecop(\bR_1)$ and $\vecop(\bR_2)$.
\item The $\ell_1$ distance (sum of absolute differences) between $\vecop(\bR_1)$ and $\vecop(\bR_2)$.
 \item $\sqrt{1-r}$, where $r$ is the correlation between the lower triangular elements of $\bR_1$ and those of $\bR_2$ (correlation of correlations); the rationale for this distance is explained in Web Appendix~\ref{supp-corcor}.
 \end{enumerate} 
We stress that (i) and (ii) are not the distances induced by the matrix 2- and 1-norms, since here we are interested in entry-wise differences as opposed to treating the matrices as operators. Distance (i) is, rather, the distance induced by the Frobenius norm, which in turn is induced by an inner product; consequently this  distance fits with the generalized true score model presented below in Section~\ref{tsm}. Since the matrices are treated here as vectors, dbICC based on distance (i) is equivalent to the I2C2 estimator of \cite{shou2013} cited at the end Section~\ref{dbrm}, although these authors focused on MRI-based images as opposed to regional connectivity matrices. 

  The dbICC estimates \eqref{dbicchat} based on distances (i)-(iii), along with 95\% bootstrap CIs, are given in the first row of Table~\ref{fctab}. While fairly consistent with the results of \cite{shou2013}, these reliabilities are very low by classical standards. 
  
    We also examined two subsets of the 333 ROIs: 41 ROIs constituting the  \emph{default mode network} of the brain \citep[DMN;][]{raichle2001}, and 39 ROIs making up the brain's visual network. Correlations among the ROIs within each of these networks tend to be high, as illustrated in Figure~\ref{brains}. Hence it comes as no surprise that dbICC values within each of these two networks, presented in the second and third rows of Table~\ref{fctab}, are markedly higher than for the complete set of ROIs. For each set of ROIs, the dbICC values are quite consistent across the three distances.
    \begin{table}[h]
	\begin{tabular}{l|l|l|l}
		\hline
		& $\ell_2$           & $\ell_1$       & $\sqrt{1-r}$              \\\hline
		All 333 ROIs & 0.378 (0.329,0.424) & 0.382 (0.335,0.426) &0.382 (0.338,0.426)  \\\hline
		Default mode network      & 0.488 (0.403,0.562) & 0.493 (0.404,0.570) &0.487 (0.414,0.555)  \\\hline
		Visual network   &0.434 (0.362,0.508)  & 0.435 (0.354,0.515) & 0.451 (0.401,0.500)\\\hline\\\\
	\end{tabular}	
\caption{Point estimates and  95\% bootstrap CIs for dbICC, based on three sets of ROIs and three distance measures.}\label{fctab}
\end{table}

\begin{figure}[!h]
	\centering
	\includegraphics[width=\textwidth]{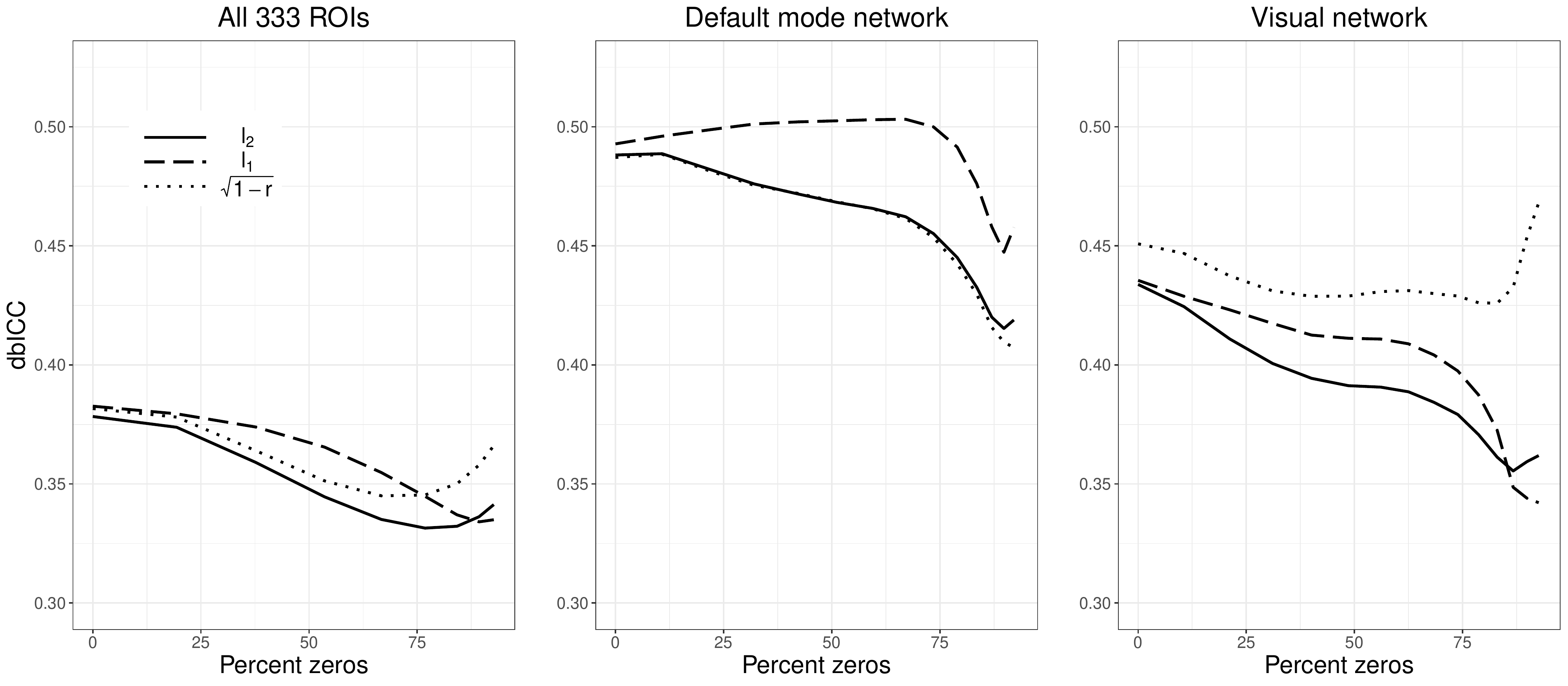}
	\caption{Estimated dbICC, for the same distances and sets of ROIs as in Table~\ref{fctab}, but with soft-thresholding of the correlation values. The horizontal axis denotes the average percentage of the correlations that are shrunk to zero, as the threshold increases.}
	\label{fig:thhd}
\end{figure}

A likely explanation for the relatively low dbICCs for the complete set of 333 ROI's is that many pairs of regions are essentially uncorrelated and thus their correlation estimates largely reflect noise. This suggests that it might be possible to boost dbICC by thresholding small correlations. Figure~\ref{fig:thhd} shows the effect on dbICC of soft-thresholding. Somewhat contrary to our expectation, soft-thresholding generally increased dbICC only slightly at best, and often decreased it.

\section{Generalizing the Spearman-Brown formula}\label{gensbf}

Is there a way to improve upon the low reliabilities found for the functional connectivity data?
A general approach to boosting reliability, suggested by classical psychometrics, is to take more measurements: for example, to average over replicates of a measure, or to increase the number of questions on a test. A well-known relation between the number of measurements and the reliability appeared in \cite{spearman1910} and, in a more familiar form, in \cite{brown1910}. In this section we extend this relation to the distance-based ICC, and in Section~\ref{sbf2} we re-examine  the fMRI data results in light of our generalization of the Spearman-Brown (SB) formula. 

\subsection{Measurement intensity and its effect on reliability}\label{measint}
The SB formula states that averaging each score over $m$ replicates  transforms the classical ICC from $\rho$ to $m\rho/[1+(m-1)\rho]$. If we let $\rho_1,\rho_m$ respectively denote the raw ICC and the ICC based on $m$ replicates, the formula can be written as $\rho_m=m\rho_1/[1+(m-1)\rho_1]$, which with some rearrangement becomes 
  \[
\frac{\rho_{m}}{1-\rho_{m}}=m\frac{\rho_1}{1-\rho_1},
\] 
or alternatively
\begin{equation}\label{sbgen}   \frac{\rho_{m}}{1-\rho_{m}}\propto m.\end{equation}
\cite{lord1968} refer to $\rho/(1-\rho)$ as the signal-to-noise ratio (SNR), and accordingly, \eqref{sbgen} may be paraphrased as: the signal-to-noise ratio is proportional to the number of measurements whose average is taken.

Averaging over $m$ real-valued measurements can be viewed as just one example of a broader notion of increasing \emph{measurement intensity} and thereby boosting reliability. Other instances of measurement intensity $m$ include:
  \begin{enumerate}[({E}1)] 
  \item An estimated covariance or correlation matrix based on a sample of $m$ multivariate observations. For functional connectivity matrices as considered above in Section~\ref{funco}, $m$ would be the number of time points recorded by fMRI.
   \label{covex}
  \item A curve estimate obtained by penalized spline smoothing with $m$ observations.\label{curvex} \end{enumerate}
Our goal in the next subsection is to derive a distance-based SB relation, i.e., an analogue of   \eqref{sbgen} in which $m$ denotes measurement intensity and $\rho_m$ is the resulting dbICC. To do this, we need a more general formulation of the true score model \eqref{xte}.

\subsection{A true score model for general Hilbert spaces}\label{tsm}
The classical setting of real-valued measures, as well as examples (E1) and (E2), can all be viewed as instances of a general setup in which the observations are of the form \eqref{xte}, but the $T_i$'s are a random sample of true scores in a Hilbert space $\cH$,  while the $\varepsilon_{ij}$'s are random measurement errors in $\cH$. We define distance in $\cH$ by $d(h_1,h_2)=\|h_1-h_2\|$, where $\|\cdot\|$ is the norm induced by the inner product on $\cH$. Define
\begin{equation}\label{delt}\Delta_T=E(\|T_{i_1}-T_{i_2}\|^2)\end{equation}
and 
  \begin{equation}\label{delv}\Delta_{\varepsilon}(m)=E_m(\|\varepsilon_{i_1j_1}-\varepsilon_{i_2j_2}\|^2),\end{equation}
for $i_1,i_2\in\{1,\ldots,I\}$ and $j_k\in\{1,\ldots,J_k\}$ for $k=1,2$, where $E_m$ denotes expectation for measurement intensity equal to $m$. Note that the measurement intensity affects only the expected distance between errors $\varepsilon_{ij}$, but not that between scores $T_i$. We make two assumptions, of which the first is implicit in  \eqref{delv}:
\begin{enumerate}[({a}1)]
\item The expectation in \eqref{delv} is the same for $i_1=i_2$ versus for $i_1\neq i_2$.
\item For all $i_1,i_2,j_1,j_2$,
\begin{equation}\label{cross0}E(\langle T_{i_1}-T_{i_2},\varepsilon_{i_1j_1}-\varepsilon_{i_2j_2}\rangle)=0.\end{equation}
\end{enumerate}
Then 
\begin{eqnarray*}\rho_m &=& 1-\frac{E(\|X_{ij_1}-X_{ij_2}\|^2)}{E(\|X_{i_1j_1}-X_{i_2j_2}\|^2)} \\
&=& 1-\frac{E(\|\varepsilon_{ij_1}-\varepsilon_{ij_2}\|^2)}{E(\|T_{i_1}+\varepsilon_{i_1j_1}-T_{i_2}-\varepsilon_{i_2j_2}\|^2)}\\&=& 
1-\frac{\Delta_{\varepsilon}(m)}{\Delta_T+\Delta_{\varepsilon}(m)}\mbox{ [by \eqref{delt},\eqref{delv},\eqref{cross0}]}\\
&=& \frac{\Delta_T}{\Delta_T+\Delta_{\varepsilon}(m)},
\end{eqnarray*}
and therefore
\begin{equation}\label{invprop}\frac{\rho_m}{1-\rho_m}=\frac{\Delta_T}{\Delta_{\varepsilon}(m)}\propto\frac{1}{\Delta_{\varepsilon}(m)}.\end{equation}

In the classical case where $X_{ij}$ is the mean of $m$ measurements, $\varepsilon_{ij}$ is the mean of $m$ independent errors with mean 0 and common variance, 
  so that \[\Delta_{\varepsilon}(m)=E(\|\varepsilon_{i_1j_1}-\varepsilon_{i_2j_2}\|^2)\propto 1/m;\] 
plugging this into \eqref{invprop} leads directly to the rearranged SB formula \eqref{sbgen}. In other cases, such as (E\ref{curvex}), $\Delta_{\varepsilon}(m)\not\propto 1/m$ and hence the generalized SB formula \eqref{invprop} does not reduce to \eqref{sbgen}.

\section{Applying the generalized SB formula to the fMRI data}\label{sbf2}

Our goal in this section is to study the implications of the generalized SB formula \eqref{invprop} for correlation matrices  such as those used in Section~\ref{funco} as measures of functional connectivity. In Section~\ref{CovE} we show that, in the simpler setting of covariance matrix estimation, the relationship between measurement intensity and reliability is essentially the same as in the classical case of scalar measures. In Sections~\ref{sbsims} and \ref{back}, 
 we investigate the extent of agreement between what is expected theoretically and what is observed with simulated and real data.

\subsection{An SB formula for covariance matrix estimation}\label{CovE}
Let $\bSigma_1,\ldots,\bSigma_I$ be a random sample of $p\times p$ covariance matrices,  and for $i\in\{1,\ldots,I\}$, let $\bS_{i1},\ldots,\bS_{iJ_i}$ be sample covariance matrices, each  based on $m$ independent and identically distributed (IID) observations $\cX_{ij1},\ldots,\cX_{ijm}$ from a $p$-variate normal distribution with covariance matrix $\bSigma_i$.  These belong to the Hilbert space $\cH$ of real symmetric $p\times p$ matrices, equipped with inner product $\langle \bA,\bB\rangle=\tr(\bA\bB^T)$; the norm induced by this inner product is the Frobenius (entry-wise $\ell_2$) norm used in the fMRI example of Section~\ref{funco}. 
 Note that here, unlike in the classical true score model, $T_i\equiv\bSigma_i$ and $\varepsilon_{ij}\equiv\bS_{ij}-\bSigma_i$ are not independent   since $\varepsilon_{ij}$ must be such that $\bS_{ij}=T_i+\varepsilon_{ij}$ is non-negative definite. 
 But as shown in the Appendix, assumptions (a1) and (a2) of Section~\ref{tsm} hold, and consequently 
\begin{equation}\label{m1}\Delta_{\varepsilon}(m)\propto \frac{1}{m-1}.\end{equation} 
Thus by \eqref{invprop}, 
\begin{equation}\label{propm1}\frac{\rho_m}{1-\rho_m}\propto m-1;\end{equation} 
this is almost exactly the classical SB relation \eqref{sbgen},  but with $m-1$ in place of $m$.

\subsection{Log-log plots with simulated data}\label{sbsims}

\begin{figure}[!h]
	\centering
	\includegraphics[width=\textwidth]{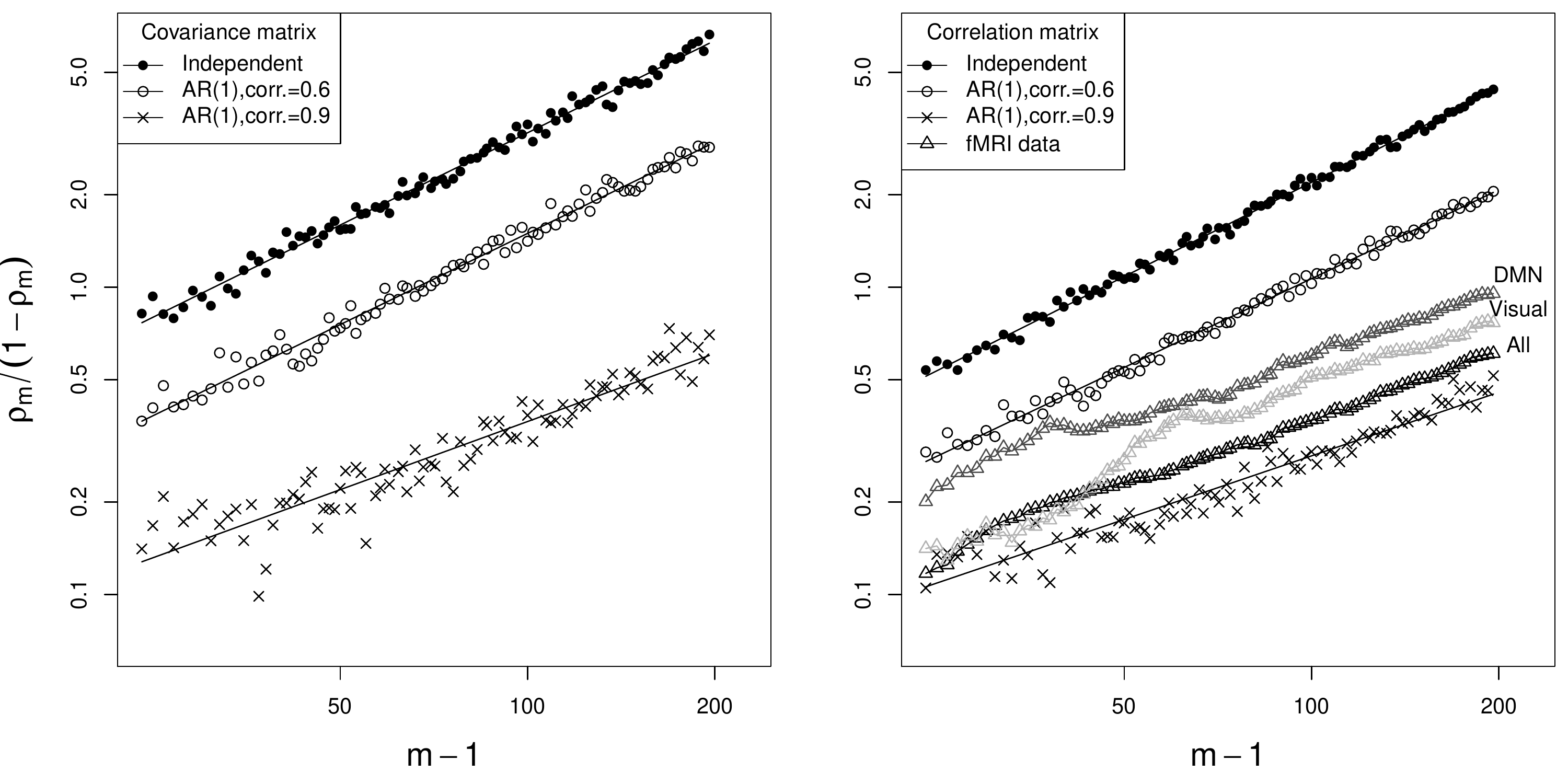}
	\caption{\emph{Left:} Effect of measurement intensity on SNR $\frac{\rho}{1-\rho}$ for covariance matrix estimation with simulated data. Both axes are plotted on the log scale since, as explained at \eqref{logm1}, this is expected to yield a linear relation with slope 1 for independent observations. \emph{Right:} Simulation results for correlation matrix estimation, along with results based on subsets of the fMRI time series.}
	\label{fig:logdepfc}
\end{figure}

Suppose that, for a given collection $\bSigma_1,\ldots,\bSigma_I$ of $p\times p$ covariance matrices, we repeatedly generate sets of sample covariances as in Section~\ref{CovE}, but with varying values of $m$, and obtain a dbICC estimate $\hat{\rho}_m$, based on the $\ell_2$ distance, for each $m$. Then the relation \eqref{propm1} suggests that the points  
\begin{equation}\label{logm1}[\log(m-1),\log\{\hat{\rho}_m/(1-\hat{\rho}_m)\}]\end{equation} should lie approximately along a line with slope 1.
To test this suggestion with simulated data resembling the fMRI data analyzed in Sections~\ref{funco} and \ref{back}, we followed the above recipe with 
\begin{itemize}
\item $I=25$, $J_i\equiv 2$ and $p=333$;
\item  $\bSigma_i$ ($i=1,\ldots,25$) taken to be the mean of the two sample covariance matrices from the $i$th participant's two fMRI scans; and 
\item a range of $m$ values from 25 to 197, approximately equally spaced on the log scale. \end{itemize}
A plot of the resulting points \eqref{logm1} appears in the left panel of Figure~\ref{fig:logdepfc} (black dots), and the best-fit line through these points has slope 0.997 with standard error 0.010, in agreement with the theoretical slope 1. 

Many aspects of the fMRI data reliability analysis in Section~\ref{funco} are not captured by the above simulation setup. Two of the most prominent disparities are that for the real data, (i)  we computed dbICC for correlation, rather than covariance, matrices, and (ii) the multivariate observations are autocorrelated rather than independent (see \cite{arbabshirani2014} and \cite{zhu2018} regarding the impact of such autocorrelation).

The simulation study was expanded to partially address these discrepancies. Using a standard implementation   \citep{barbosa2012} for vector autoregressive models of order 1 \citep[VAR(1);][]{lutkepohl2005}, we conducted further simulations in which the $j$th multivariate time series for the $i$th individual was given by $\bx^{(ij)}_t=\phi\bx^{(ij)}_{t-1}+\bu^{(ij)}_t$ ($t=2,\ldots,m$), with independent innovations $\bu^{(ij)}_t$ having zero mean and $333\times 333$ covariance matrix $\bSigma_i$. The lag-1 autocorrelation $\phi$ was set to the values 0.6 and 0.9, which are near the low and high ends of the range of AR(1)-model-based estimates for individual ROIs in our fMRI data. The resulting points  \eqref{logm1}, with $\hat{\rho}_m$ derived from sample covariance matrices, are displayed in the left panel of Figure~\ref{fig:logdepfc}. The right panel is analogous, but here $\hat{\rho}_m$ is derived from sample correlation matrices. A comparison of the two panels indicates that, for given autocorrelation settings, both the estimated SNR $\hat{\rho}_m/(1-\hat{\rho}_m)$ and its dependence on $m$ are very similar for covariance versus correlation matrix estimation. Autocorrelation is seen to reduce reliability and thus to shift the SNR markedly downward. Moreover, autocorrelation seems to attenuate the linear relationship between $m$ and SNR: whereas in the IID setting the slope is 1.018 for the sample correlation matrix, again very close to the theoretical value 1, the slopes are smaller with autocorrelation 0.6 (0.986 for covariance, 0.960 for correlation) and even smaller for autocorrelation 0.9 (0.736 for covariance, 0.687 for correlation). In Web Appendix~\ref{supp-table33} we present  plots that are analogous to Figure~\ref{fig:logdepfc}, but based on the $\ell_1$ and $\sqrt{1-r}$ distances, and we report the intercepts and slopes of the best-fit lines for all cases.

\subsection{Reliability based on subsets of the fMRI time series}\label{back}
Next we constructed  log-log plots as above but based on subsets of the real fMRI time series of Section~\ref{funco} rather than on simulated data. For values of $m$ ranging from 25 to the full time series length 197, we took the middle $m$ observations from each of the fMRI time series, and thus computed correlation matrices $\bR_{ij}$ ($i=1,\ldots,25;j=1,2$) using the same three sets of ROIs as in Section~\ref{funco}: all 333 ROIs proposed by \cite{gordon2016}, the default mode network, and the visual network.  Log-log plots for the resulting dbICC values $\hat{\rho}_m$ appear in the right panel of Figure~\ref{fig:logdepfc}. For smaller $m$ these plots are quite nonlinear and distinct from each other, but for $m>100$, they each appear to stabilize with a linear pattern that is roughly parallel to the best-fit line for the simulations with lag-1 autocorrelation 0.9. 

This degree of agreement with the simulation results of Section~\ref{sbsims} is probably as much as can be expected, given the significant discrepancies between the settings of the simulated- and real-data analyses, which include the following. (i) The simulations for different $m$ are independent, whereas with the real data, for increasing $m$ we consider a nested sequence of increasingly large subsets of the same time time series. (ii) The real time series may not be multivariate normal and presumably have more complex patterns of autocorrelations and cross-correlations than the simulated data. 

At any rate, it seems clear that the theoretical log-log plot slope of 1 cannot be expected to characterize the reliability improvement attainable via longer fMRI time series. Our results offer hope that a slope around 0.7 might be attained, but at least two further caveats are in order. One is that we cannot extrapolate beyond $m=197$, the full time series length for our data.
A second, subtler caveat concernes the true score model \eqref{xte}, in the specific form outlined in Section~\ref{CovE}. That model assumes that for each $i$, the two sample covariance matrices $\bS_{i1},\bS_{i2}$ are estimates of a common true covariance $\bSigma_i$. But if in fact the underlying covariance matrix differs between the two fMRI scans for at least some of the participants, this is an additional source of within-subject distance that is not removed by increasing the time series length $m$, and thus $\log[\hat{\rho}_m/(1-\hat{\rho}_m)]$ may tend to level off rather than increasing linearly with  $\log(m-1)$. In summary, 
  while longer fMRI scans might make correlation matrices more reliable as measures of functional connectivity,
  the improvement would likely be less dramatic than the results reported here might lead us to expect. 


\section{Further application and extension of the SB formula}\label{mgd}
Log-log plots like those  in Figure~\ref{fig:logdepfc} are a broadly applicable tool for examining the relationship between measurement intensity $m$ and reliability. As discussed in Web Appendix~\ref{supp-curvest}, for penalized spline smoothing (example (E\ref{curvex}) of Section~\ref{measint}), $\Delta_{\varepsilon}(m)\propto m^{-8/9}$. Thus, arguing as in Section~\ref{sbsims}, a linear model fit to the points  $[\log(m),\log\{\hat{\rho}_m/(1-\hat{\rho}_m)\}]$ should have slope $\frac{8}{9}$, a prediction that is borne out with simulated data.

Some distances, such as the dynamic time warping distance between signatures considered in Web Appendix~\ref{supp-sigdat}, do not arise from the true score model \eqref{xte}, even in the generalized (Hilbert space-valued) form of Section~\ref{tsm}. Whether or not the true score model applies, the dbICC \eqref{dbicc} satisfies
\begin{equation}\label{dbsat}\frac{\rho}{1-\rho}=\frac{\msd_b-\msd_w}{\msd_w}.\end{equation}
The key to the derivation of  \eqref{invprop} is simply that, by \eqref{delt}--\eqref{cross0}, 
\begin{enumerate}[(i)]
\item $\msd_w=\msd_w(m)=\Delta_{\varepsilon}(m)$,
\item $\msd_b-\msd_w=\Delta_T$, which does not depend on $m$. 
\end{enumerate}
The same argument works more generally (i.e., not only in Hilbert spaces): as long as $\msd_w$ can be written as a function of $m$ whereas $\msd_b-\msd_w$ does not change with $m$, it follows from \eqref{dbsat} that 
\begin{equation}\label{ip2}\frac{\rho_m}{1-\rho_m}\propto\frac{1}{\msd_w(m)},\end{equation}
generalizing \eqref{invprop}, which is itself a generalization of \eqref{sbgen}.

  Log-log plots might be used in this more general setting to estimate the effect of measurement intensity $m$ on $\rho_m$, as opposed to confirming a theoretical relationship.
   By \eqref{ip2}, if it is expected that $\msd_w(m)\propto m^{-\beta}$ for some  unknown $\beta$, then we can regress values of $\log\frac{\hat{\rho}_m}{1-\hat{\rho}_m}$ on the corresponding values of $\log(m)$, and the resulting slope serves as an estimate of $\beta$. A similar approach is used to estimate the Hurst exponent of a long memory process \citep{beran1994}.


\section{Discussion}
\label{s:discuss}
In this paper we have redefined the intraclass correlation coefficient in terms of distances, and thereby extended this reliability index to arbitrary data objects for which a distance is defined. The proposed distance-based ICC leads to two extensions of the SB formula, namely \eqref{invprop} for Hilbert space-valued data including covariance matrices, and \eqref{ip2} for more general data objects.

In  an early paper on extending the ICC to multivariate data, \cite{fleiss1966} wrote that a classical (univariate) ICC value less than about 0.70 ``is, for most purposes, taken to indicate insufficient reliability.'' The much lower dbICC values that we report for functional connectivity data, along with similar results reported by others \citep[e.g.,][]{shou2013}, are a sobering indication that in some cases, as technology has advanced, the reliability of complex new measures has retreated. This might help to explain the recently-much-discussed difficulties surrounding scientific reproducibility, a desideratum that is closely related to reliability \citep{yu2013}.

While our presentation has focused on test-retest data, the dbICC might also be applied to assess the reliability  of results obtained by algorithms, such as bootstrapping, that have a stochastic component \citep[cf.][]{philipp2018}.

Whereas we have developed a distance-based analogue of the \emph{intra}class correlation coefficient, the distance correlation of \cite{szekely2007} is comparable to \emph{inter}class correlation coefficients. Extending ideas from distance correlation research to the intraclass setting may be an interesting avenue for future work.

A package for R \citep{R} implementing the methods of this paper is available at \url{https://github.com/wtagr/dbicc}.


\backmatter


\section*{Acknowledgements}
The authors thank the Co-Editor, Mark Brewer, the Associate Editor and the reviewers for very helpful and thoughtful feedback. Thanks are due as well to Eva Petkova and Don Klein for inspiring this work, by calling attention to the need for reliable measurement in the early days of resting-state fMRI connectivity research. The work of M.~Xu and P.~T. Reiss was supported by Israel Science Foundation grants 1777/16 and 1076/19. The work of I.~Cribben was supported by Natural Sciences and Engineering Research Council (Canada) grant RGPIN-2018-06638 and the Xerox Faculty Fellowship, Alberta School of Business. 

\section*{Data availability statement}
The data that support the findings of this study are available at \url{https://github.com/wtagr/dbicc}. These data were derived from the public-domain NYU CSC TestRetest resource at \url{http://www.nitrc.org/projects/nyu\_trt}.

\section*{Supplementary Materials}

Web Appendix~\ref{supp-logR}, referenced in Section~\ref{s:intro}, Web Appendix~\ref{supp-corcor}, referenced in Section~\ref{funco}, Web Appendix~\ref{supp-table33}, referenced in Section~\ref{sbsims}, and Web Appendices~\ref{supp-curvest} and  \ref{supp-sigdat}, referenced in Section~\ref{mgd},   are available with
this paper at the Biometrics website on Wiley Online
Library.\vspace*{-8pt}


%
\bibliographystyle{biom} 
\bibliography{icc}

\appendix

\section{}
\subsection{fMRI data description and preprocessing}\label{pipe}
The resting-state fMRI data set, downloaded from \url{http://www.nitrc.org/projects/nyu\_trt}, includes 25 participants (mean age 29.44 $\pm$ 8.64, 10 males) scanned at  New York University. A Siemens Allegra 3.0-Tesla scanner was used to obtain three resting-state scans for each participant, though for this analysis, we  considered only the second and third scans, which were less than one hour apart. Each scan consisted of 197 contiguous EPI functional volumes with time repetition (TR) = 2000 ms; time echo (TE) = 25 ms; flip angle (FA) = $90^{\circ}$; 39 number of slices, matrix = $64\times64$; field of view (FOV) = 192 mm; voxel size $3\times3\times3$ mm$^3$. During each scan, the participants were asked to relax and remain still with eyes open. For spatial normalization and localization, a high-resolution T1-weighted magnetization prepared gradient echo sequence was obtained (MPRAGE, TR = 2500 ms; TE = 4.35 ms; inversion time = 900 ms; FA = $8^\circ$, number of slices = 176; FOV = 256 mm).

The data were preprocessed using  the {\tt FSL} (\url{http://www.fmrib.ox.ac.uk}) and {\tt AFNI} (\url{http://afni.nimh.nih.gov/afni}) software packages. The images were (i) motion corrected using {\tt FSL}'s {\tt mcflirt} (rigid body transform; cost function normalized correlation; reference volume the middle volume) and then (ii) normalized into the Montreal Neurological Institute space using {\tt FSL}'s {\tt flirt} (affine transform; cost function mutual information). (iii) {\tt FSL}'s {\tt fast} was then used to obtain a probabilistic segmentation of the brain to acquire white matter and cerebrospinal fluid (CSF) probabilistic maps, thresholded at 0.99. (iv) {\tt AFNI}'s {\tt 3dDetrend} was then used to  remove the nuisance signals, namely the six motion parameters, white matter and CSF signals, and the global signal. (v) Finally, using {\tt FSL}'s {\tt fslmaths}, the volumes were spatially smoothed using a Gaussian kernel with FWHM = 6mm.

The ROIs for our connectivity analysis are derived from the work of \cite{gordon2016}, who parcellated the cortical surface into 333 areas within which homogeneous connectivity patterns are observed. Time courses for these 333 ROIs were obtained for each subject by averaging over all of the voxels within each region. Each regional time course
was then detrended and standardized to unit variance, and then we applied a 4th-order Butterworth filter with passband 0.01--0.10 Hertz.

\subsection{(a1), (a2) and $\Delta_{\varepsilon}(m)$ for sample covariance matrices}\label{edcov}
Sample covariance matrices of multivariate normal samples are a special case of the true score model of Section~\ref{tsm} in which, for each $i$, $T_i=\bSigma_i$, a  $p\times p$ covariance matrix, and for each $i,j$, 
\begin{equation}\label{vij}\varepsilon_{ij}=\bS_{ij}-\bSigma_i,\end{equation} where $\bS_{ij}$ is the sample covariance matrix of an IID random sample $\cX_{ij1},\ldots,\cX_{ijm}\sim N_p(\bzero,\bSigma_i)$. Here we verify assumptions (a1) and (a2) of Section~\ref{tsm} for this case, and derive expression \eqref{m1} for $\Delta_{\varepsilon}(m)$.

By \eqref{vij}, $\varepsilon_{i_1j_1},\varepsilon_{i_2j_2}$ in \eqref{delv} are independent mean-zero matrices, implying that \begin{eqnarray*}\Delta_{\varepsilon}(m)&=&E\left[\tr\{(\varepsilon_{i_1j_1}-\varepsilon_{i_2j_2})^2\}\right]\\&=&E[\tr(\varepsilon_{i_1j_1}^2)]+E[\tr(\varepsilon_{i_2j_2}^2)]-2E[\tr(\varepsilon_{i_1j_1}\varepsilon_{i_2j_2})].\end{eqnarray*}
For $i_1\neq i_2$, $E[\tr(\varepsilon_{i_1j_1}\varepsilon_{i_2j_2})]=0$ since $\varepsilon_{i_1j_1},\varepsilon_{i_2j_2}$ are independent mean-zero matrices. On the other hand, if $i_1=i_2=i$ then $\varepsilon_{i_1j_1},\varepsilon_{i_2j_2}$ are independent and of mean zero, conditionally on $\bSigma_i$, and thus again 
\[E[\tr(\varepsilon_{i_1j_1}\varepsilon_{i_2j_2})]=E[E\{\tr(\varepsilon_{i_1j_1}\varepsilon_{i_2j_2})|\bSigma_i\}]=0.\]
Hence the expectation defining $\Delta_{\varepsilon}(m)$ does not depend on whether or not $i_1=i_2$, i.e., (a1) holds; and
\begin{equation}\Delta_{\varepsilon}(m)=2E[\tr(\varepsilon_{ij}^2)],\label{trej}\end{equation}
for $\varepsilon_{ij}$ as in \eqref{vij}. 

For (a2), it suffices to show that $E[\tr\{\bSigma_{i1}(\varepsilon_{i_1j_1}-\varepsilon_{i_2j_2})\}]=0$. This follows since
\[E[\tr(\bSigma_{i_1}\varepsilon_{i_1j_1})]=E[E\{\tr(\bSigma_{i1}\varepsilon_{i_1j_1})|\bSigma_{i_1}\}]=0,\]
while $E[\tr(\bSigma_{i1}\varepsilon_{i_2j_2})]=0$ since $\varepsilon_{i_2j_2}$ is independent of $\bSigma_{i1}$ and of mean zero.

By a standard result in multivariate analysis, conditionally on $\bSigma_i$, $(m-1)\bS_{ij}$ has a Wishart$(\bSigma_i)$ distribution with $m-1$ 
	degrees of freedom; thus by Theorem~2.2.6 of \cite{fujikoshi2010}, 
	\[ E[\tr(\bS_{ij}^2)|\bSigma_i]=\frac{1}{m-1}\left[(\tr\bSigma_i)^2+m
	\tr(\bSigma_i^2)\right]  \quad\mbox{and}\quad E[\tr(\bS_{ij}\bSigma_i)|\bSigma_i]=\tr(\bSigma_i^2).\]
These results lead to
\begin{eqnarray*}E[\tr(\varepsilon_{ij}^2)|\bSigma_i] &=& E[\tr\{(\bS_{ij}-\bSigma_i)^2\}|\bSigma_i]  \\
&=& \frac{1}{m-1}[(\tr\bSigma_i)^2+\tr(\bSigma_i^2)].\end{eqnarray*}
Combining this with \eqref{trej} gives \[\Delta_{\varepsilon}(m)=\frac{2}{m-1}E[(\tr\bSigma_i)^2+\tr(\bSigma_i^2)],\]
where the expectation is with respect to the distribution of the true covariance matrices $\bSigma_i$. This confirms \eqref{m1}.

\label{lastpage}

\end{document}

%% file: bold.tex
\newcommand{\bu}{\mbox{\boldmath $u$}}

\newcommand{\bx}{\mbox{\boldmath $x$}}

\newcommand{\bA}{\mbox{\boldmath $A$}}
\newcommand{\bB}{\mbox{\boldmath $B$}}

\newcommand{\bI}{\mbox{\boldmath $I$}}

\newcommand{\bR}{\mbox{\boldmath $R$}}
\newcommand{\bS}{\mbox{\boldmath $S$}}

\newcommand{\bzero}{\mbox{\bf 0}}

\newcommand{\bSigma}{\mbox{\boldmath $\Sigma$}}

\newcommand{\cH}{\mathcal{H}}

\newcommand{\cX}{\mathcal{X}}